\documentclass[preprint,showpacs,preprintnumbers,amsmath,amssymb]{revtex4}
\usepackage{graphicx,epsfig}
\usepackage{dcolumn}
\usepackage{bm}
\begin{document}
\title{ Dynamical transition in translational and rotational dynamics of water in the grooves of DNA duplex at low temperature}
\author{Debasmita Biswal, Biman Jana, Subrata Pal and Biman Bagchi}
\email{bbagchi@sscu.iisc.ernet.in}
\affiliation{Solid State and Structural Chemistry Unit, Indian Institute of 
Science, Bangalore 560012, India}

\begin{abstract}
We have simulated structure and dynamics of water in the grooves of a DNA duplex 
using moleculear dynamics simulations. We find signatures of a dynamical transition 
in both translational and orientational dynamics of water molecules in both the major 
and the minor grooves of a DNA duplex. The transition occurs at a slightly higher 
temperature ($T_{GL} \approx $ 255 K) than the temperature ($T_L\approx$ 247 K) 
where the bulk water is conjectured to undergo a dynamical transition. Groove water, 
however, exhibits markedly different temperature dependence of its properties from the 
bulk. Entropy calculations reveal that the minor groove water is ordered even at room 
temperature and the transition at $T \approx$ 255 K can be characterized as 
a {\em strong-to-strong} dynamical transition. The low temperature water is characterized 
by pronounced tetrahedral order, as manifested in the sharp rise near $109^{\circ}$ 
in the O-O-O angle distribution. We find that Adams-Gibbs relation between 
configurational entropy and translational diffusion holds quite well when the two 
quantities are plotted together in a master plot for different region of aqueous 
DNA duplex (bulk, major and minor grooves) at different temperatures. The activation 
energy for the transfer of water molecules between different regions of DNA is found to 
be independent of temperature. 
\end{abstract}

\pacs{64.70.Ja, 61.20.Ja, 61.20.Gy}

\maketitle

\section{Introduction}
Water in the natural world is often found under constrained and/or restricted 
environments, in the hydration layer of proteins and micelles, within reverse 
micelles and microemulsions, in the grooves of DNA duplex, within biological 
cells, to name a few. Properties of water under such constrained conditions 
can be quite different from those of bulk, neat water~\cite{bb_chemrev05}. 
However, it is likely that even under such restricted conditions water 
retains some of its unique properties. Study of these unique properties 
of water, especially in the hydration layer of biomolecules, particularly 
of proteins~\cite{bb_chemrev05,cheng_rossky_nature98,tarek_tobias_prl02,lehninger_book,
tarek_tobias_biophys00,murarka_gordon08} 
and DNA~\cite{bb_chemrev05,pal_jcp06,jana_entropy,berg_Physchemphys,andreatta_JACS_2006, 
sen_JACS_2008, jayram_ARBBS_96}, has been a subject of great interest in recent 
times. Hydration layer not only provides 
the stability of the structure of the biomolecules, but also plays a 
critical role in the dynamic control of biological activity. The 
intercalation of anti-tumor drugs, such as daunomycin, into DNA involves 
active participation of water molecules in the 
grooves~\cite{qu_pnas01,arnab_jacs08}.

The low temperature (near 200 K) ``glasslike'' transition of hydrated 
protein has drawn a great deal of attention in both experimental and 
computer simulation studies~\cite{ringe_biophyschem03,hartmann_pnas82,
wand_nature01,rasmussen_nature92,doster_nature89}. Above this transition 
temperature proteins exhibit diffusive motion and below this temperature 
the proteins are trapped in localized harmonic modes. An important issue 
in recent times is to determine the effects of hydration water on this 
dynamical transition~\cite{vitkup_nsb00,tarek_tobias_bj00,zanotti_bj99,
sololov_jcp99,nilsson_pnas96}. Recent studies have shown that dynamics of 
water in the hydration layer of a protein also exhibits strong temperature 
dependence around the same temperature and it seems to undergo a 
fragile-to-strong transition which preempts an otherwise expected glass 
transition at a lower temperature~\cite{nilsson_pnas96,lagi_jpcb08,
cheng_jcp06,mellamace_jcp07}. 

Study of DNA hydration layer has recently indicated interesting dynamical 
behavior of water in the grooves~\cite{pal_jcp06,jana_entropy}. Several recent studies 
have discussed about the origin of the slow component of the solvation dynamics in DNA 
hydration layer~\cite{sen_JACS_2008}. However, a detail discussion of this upcoming 
issue is beyond the scope of this paper. 

A recent computer simulation study by Stanley and co-workers has shown that the 
liquid-liquid (L-L) transition in water 
induces a {\em dynamic transition} in DNA which has striking 
resemblance with that of liquid to glass transition~\cite{stanley_prl06}. 
{\em This study, however, did not explore the dynamics of water in the grooves of DNA}. 

There are many questions that have remained unanswered 
regarding dynamics of groove water at low temperature. For example, is there 
a dynamic transition in the grooves of DNA near the L-L transition of 
bulk water? Does it in any way resemble the one in protein hydration layer?  
Note that the remarkable properties of bulk water have recently been 
attributed to a {\em highly interesting L-L transition} at 
around $T_L \approx$ 247 K, that is, only 26$^o$C below the freezing 
temperature~\cite{harrington,mayer_jpcb99,xu_pnas05}. The effects of the 
bulk water L-L transition on groove water  dynamics have not yet been 
investigated.

In this article we report our finding that water in the {\em grooves of a 
DNA duplex} shows a dynamical transition at a temperature ($T_{GL}$) 
slightly higher than the temperature ($T_L$) where the bulk water undergoes 
the L-L transition. However, the nature and manifestation of the transition 
in the grooves are quite different from that in the bulk.

\section{System and Simulation Details}
The system we studied consists of a {\em Dickerson dodecamer} DNA 
duplex (CGCGAATTCGCG)~\cite{drew_pnas81} solvated in 1565 TIP5P water 
molecules~\cite{tip5p}. We have studied the DNA-water system at constant 
pressure $P$ = 1 atm, at several constant temperatures (NPT ensemble) in 
a simulation box with periodic boundary condition. The molecular dynamics 
simulations of this aqueous DNA system were performed using the AMBER Force 
Field~\cite{amber}.

We have identified the groove water by using the following procedure. 
We have calculated the radial distribution function (g(r)) of water 
molecules in the system from the major and minor groove atoms. On the 
basis of this g(r), a cut-off distance of 3.5\AA~(the first minima of 
g(r)) from the groove atoms is used for the selection. For bulk water 
analysis, we have considered those water molecules which are beyond 
15\AA~from any DNA atoms. We have checked that at 15~\AA~away, 
water indeed regains bulk-like behavior.

\section{Results and Discussions}
\subsection{Mean square fluctuation of DNA and translational diffusivity of water} 
We first report the calculated mean square atomic fluctuation 
$\langle \rm X^2 \rangle$ of the DNA atoms starting from 300 K down to 
210 K in order to characterize the macromolecular ``glass'' transition 
temperature ($T_{DNA}$). Left panel of {\bf Figure 1} displays 
the same. We find that the mean square fluctuation (MSF) of DNA slows 
down dramatically around 247 K and continues to remain slow for the lower 
temperatures. The onset of the change in slope (near $T_{DNA} \approx$ 247 K) 
of MSF indicates a macromolecular dynamic transition. Right panel of 
{\bf Figure 1} shows temperature dependence of 
the diffusivity for all the water molecules in the system. It shows a 
crossover around the same temperature ($T_L\approx$ 247 K) from a high 
temperature power law form to a low temperature Arrhenius form. From the power 
law fit to the high temperature region we get a glass transition 
temperature of 231 K which is in agreement with the earlier simulation 
study by Stanley and co-workers~\cite{stanley_prl06}. 

\begin{figure}
\begin{center}
\epsfig{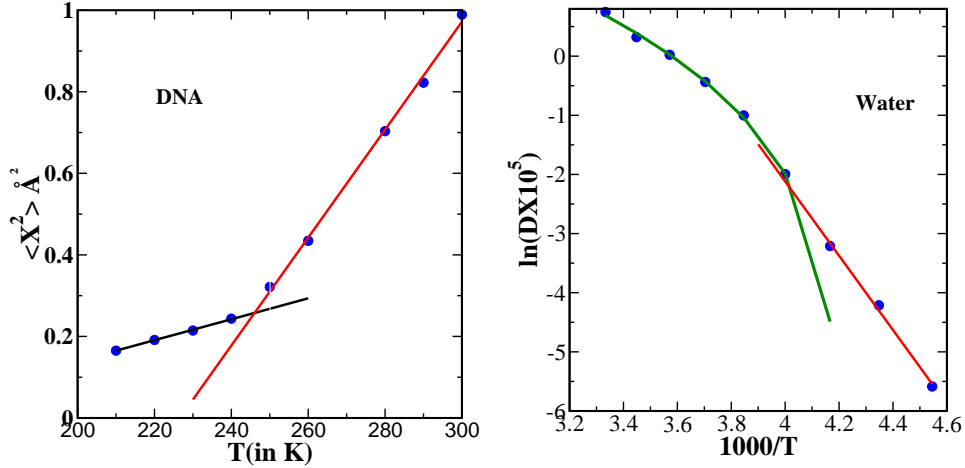}
\end{center}
\caption{Mean square fluctuation (MSF) of DNA duplex (left panel) and
diffusivity (right panel) of the oxygen atoms of all water in the system. In the left (DNA)
panel, MSF of DNA shows a dynamic transition at $T\approx$ 247 K. In the right (water) panel,
water shows dynamical crossover around same temperature from a high temperature power law
behaviour (cyan) to a low temperature Arrhenius behaviour (red).}
\end{figure}

\subsection{Intermediate scattering function} 
We next discuss the self intermediate scattering function (ISF) of the 
oxygen atoms of the water molecules in the grooves of DNA for a set of 
temperatures starting from 300 K to 210 K. The self intermediate scattering 
function is defined as 
\begin{equation}
\label{eq:fskt}
F_S(k,t)= \langle exp(-ik\cdot(r(t)-r(0)))=
\left \langle \frac{sin|k||r(t)-r(0)|}{|k||r(t)-r(0)|}\right \rangle,
\end{equation}
where $k$ is the wave vector and $r(t)$ is the position of the oxygen atom 
of the water molecules. The $\vert k \vert$ value taken here 
is 2.5\AA$^{-1}$.~The 
translational relaxation time ($\tau_T$) is obtained by fitting the two 
step relaxation of ISF at different temperatures using Relaxing Cage 
Model (RCM)~\cite{chen_pre99}. The fitting equation used here is given by
\begin{equation}
\label{eq:fskt_fit}
F_S(k,t)=[1-A(k)]e^{-(t/\tau_{S})^2} + A(k)e^{-(t/\tau_{T})^\beta}
\end{equation}
Here A(k) is Debye-Waller factor, $\tau_T$ being the translational relaxation 
time and $\beta$ is the stretched exponent. 

{\bf Figure 2(a)} shows ISF of 
oxygen atom for the water molecules in bulk, major groove and minor groove 
at 300 K and 260 K. It is evident from both the figures that water molecules 
in both the major and the minor groove tend to behave like a liquid at a 
temperature lower than the bulk. The behavior is more prominent for water 
molecules in the minor groove. This can be ascribed to the fact that 
translational motion of water molecules in the minor groove is more 
constrained owing to the more ordered structure in the minor groove than 
water molecules in the major groove of DNA. Water molecules in major groove are, 
in turn, translationally more constrained than bulk water. {\bf Figure 2(b)} 
shows the temperature dependence of $\tau_T$ for water molecules in bulk, 
major and minor grooves. For both bulk and major groove water the temperature 
dependence at high temperature region can be fitted to the 
Vogel-Fulcher-Tammann (VFT) law, $\tau_T = \tau_0~exp[DT_0/(T-T_0)]$, where $D$ 
is a constant measuring fragility and $T_0$ is ideal glass transition 
temperature at which relaxation time diverges. In reality, however, the 
divergence is avoided as below a certain characteristic temperature, the 
functional dependence of relaxation time switches over to an Arrhenius 
form which is a signature of a strong liquid. The crossover temperatures for 
bulk water and major groove water are found to be $T_L\approx$ 247 K and 
$T_{GL}\approx$ 255 K, respectively. The dynamical transitions are of 
fragile-to-strong type, {\em although the fragility of major groove 
water is smaller of the two}. 
\begin{figure}
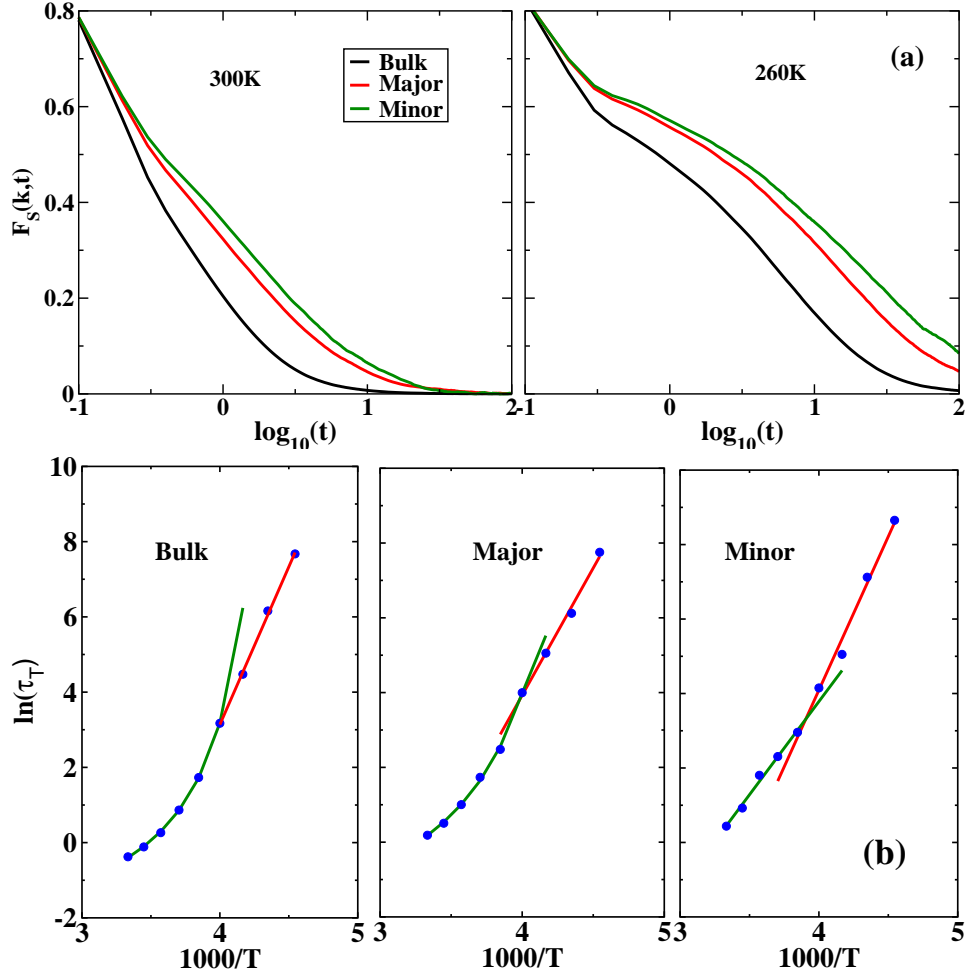

\epsfig{file=Bbagchi_jpc_fig2a.eps,width=5in,angle=0}
\epsfig{file=Bbagchi_jpc_fig2b.eps,width=5in,angle=0}
\caption{(a) Intermediate scattering function ($F_S(k,t)$) of the oxygen
atoms of the water molecules in bulk, major and minor grooves of DNA duplex
at two different temperatures, T = 300 K (left panel) and T = 260 K (right
panel) for  $\vert k \vert$ = 2.5\AA$^{-1}$.(b) Translational relaxation time
($\tau_T$) for bulk (left panel), major groove (middle panel) and minor
groove (right panel) water. Bulk and major groove water show dynamical
crossover between high temperature VFT behaviour (cyan) and low temperature 
Arrhenius behaviour (red). Minor
groove water shows transition between two Arrhenius behaviours.}
\end{figure}

The minor groove water molecules however show a remarkably different 
translational dynamics. Minor groove water does not show any signature 
of a fragile liquid in the temperature range studied. Instead, temperature 
dependence of translational relaxation time for minor groove water fits 
well into {\em two Arrhenius forms} of different slopes with a cross-over 
temperature around $T_{GL}\approx$ 255 K (same as major groove water). This 
can be understood in the context of difference in the structure of hydration 
layer in the grooves of DNA. Hydration in the minor groove is more extensive, 
regular with a {\em zig-zag} spine of first and second shell of hydration 
where as hydration in major groove is restricted to a single layer of water 
molecule~\cite{beveridge_pnas88}. Water molecules in minor groove are thus 
more structured in comparison with major groove water which results in a 
strong liquid type of behavior for water molecules in minor groove even in 
the high temperature region. This explains why in contrast to bulk and major 
groove water, minor groove water shows a {\em strong-to-strong} type of 
dynamical transition. 

\subsection{Orientational dynamics}
We next analyze orientational (dipole-dipole) time correlation 
function (TCF) of water molecules in the different regions of aqueous 
DNA and the TCF calculated is defined as
\begin{equation}
\label{eq:dipdip}
C_\mu(t) = \frac{\langle \mu(0)\cdot\mu(t)\rangle}
                {\langle \mu(0)\cdot\mu(0)\rangle}
\end{equation}
where $\mu(t)$ is the dipole moment unit vector of the water molecule at 
time t and the angular bracket corresponds the ensemble averaging. 
\begin{figure}
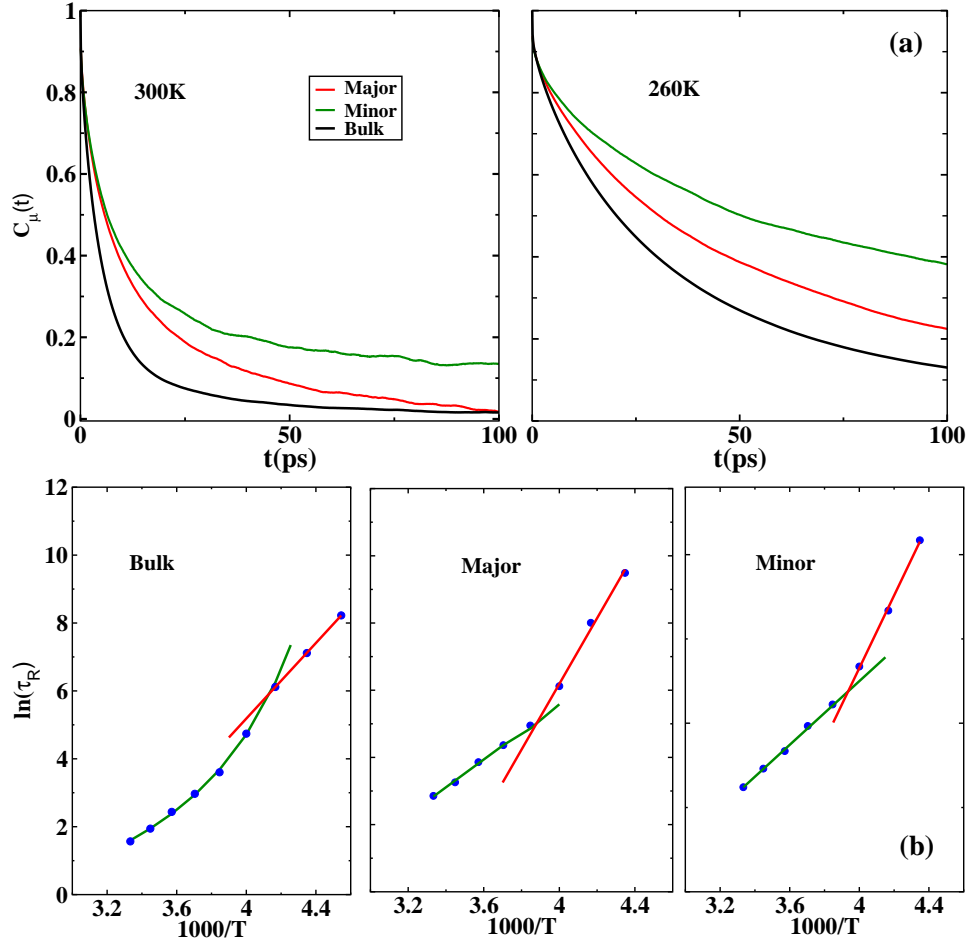

\epsfig{file=Bbagchi_jpc_fig3a.eps,width=5in,angle=0}
\epsfig{file=Bbagchi_jpc_fig3b.eps,width=5in,angle=0}
\caption{ (a) Dipole-dipole time correlation function ($C_\mu(t)$) of
water molecules in bulk, major groove and minor groove of DNA duplex at
two different temperatures, T = 300 K (left panel) and T = 260 K (right
panel). (b) Rotational relaxation time ($\tau_R$) for bulk (left panel), major
groove (middle panel) and minor groove (right panel) water. Bulk water
shows a crossover between high temperature VFT behaviour (cyan) and low temperature 
Arrhenius behaviour (red).
Major groove and minor groove water show transition between two Arrhenius
behaviour.} 
\end{figure}

{\bf Figure 3(a)} displays $C_\mu(t)$ for water molecules in bulk, major 
and minor grooves at 300 K and 260 K, respectively. Similar to the 
translational motion, rotation of the minor groove water molecules is found 
to be the most constrained. {\bf Figure 3(b)} shows the temperature dependence 
of orientational relaxation time ($\tau_R$) as obtained from stretched 
exponential fitting at long time of dipole-dipole TCF for bulk, major and 
minor groove water. Bulk water shows a fragile-to-strong 
transition around the same temperature ($T_L\approx$ 247K) as observed 
for translational relaxation time. However, unlike translational relaxation, 
orientational relaxation shows a transition between two Arrhenius forms 
of different slopes with a crossover temperature $T_{GL}\approx$ 255 K 
for {\em both major and minor groove water}. Strong-to-strong transition 
observed in the minor groove (both translational and orientational dynamics) 
can be attributed to the effect of confinement in the minor groove (higher 
depth and lower width). It is known that a confined liquid is comparatively 
less fragile than in the bulk~\cite{depablo_prl06,soles_prl02} and  this 
eventually gives rise to a Arrhenius temperature dependence of the 
relaxation times (signature of strong liquids) {\em even at the higher 
temperature region}. The reason for the different behavior of major 
groove water is thus probably due to the fact that rotation probes local 
environment more faithfully than translation.

\section{Microscopic characterization: O-O-O angle distribution}
To understand how the structure of groove water changes across the dynamical transition,
we have calculated the O-O-O angle distribution inside the first coordination shell of
a water molecule. Angle distributions at three different temperatures (300K, 250K 
and 230K) for groove water molecules are displayed in {\bf Figure 4}. At all the temperatures,
the distribution has a two peak character. While the peak at lower angle is the signature 
of the presence of interstitial water molecules inside the first hydration shell, higher 
angle peak characterizes the degree of tetradehrality present. As it is evident from this 
{\bf Figure}, with decreasing temperature the degree of tetrahedrality increases (higher 
angle peak height increases) with the removal of interstitial water molecules 
(lower angle peak height decreases) from the first hydration shell. Structural change of 
this kind with decreasing temperature and increasing order (as discussed further below) 
 responsible for the dynamical transition 
for groove water molecules. 

Bulk water also exhibits a dynamical transition near 250K. The signatures are, however, 
weaker in the case of bulk water than what are observed in the grooves. The role of 
confinement  in fostering the transition of tetraedral water can be understood in the 
following fashion. In the confined state, water molecules gain in energy but lose 
entropy (see next section). The tetrahedrally coordinated water is a low entropy state 
of the system. Confinement thus favours the crossover/transition to the tetrahedral 
state. 

\begin{figure}
\epsfig{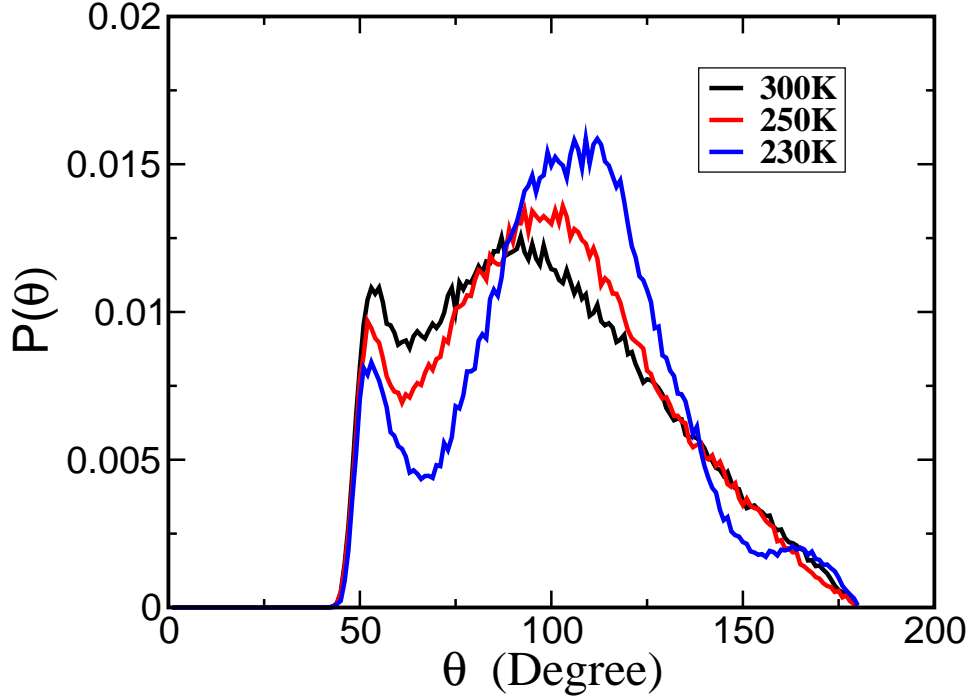}
\caption{ O-O-O angle distribution of the groove water molecules inside the 
first coordination shell at 300K, 250K and 230K . Note the decrease of 
lower angle (interstitial) peak and increase of higher angle (degree of 
tetrahedrality) peak height with decreasing temperature.}
\end{figure}

\section{Entropy calculation}
In order to understand the origin of the large observed differences 
between the dynamics of water molecules in the minor groove and in the 
bulk, we have calculated the entropy of water molecules in the respective 
regions~\cite{jana_entropy,goddard_jcp03} at two different temperatures 
(300 K and 280 K). In both the temperatures, minor groove water 
molecules have substantially lower entropy than bulk. At 300 K the difference 
is $\sim$ 60\% of the latent heat of fusion of bulk water. Entropy is 
usually found to be closely correlated with diffusion coefficient and 
structural relaxation time, in {\bf Figure 5}, we show the correlation 
between $TS_{Conf}$ and translational diffusivity and show that the 
{\em Adam-Gibbs relation} remains valid for the different regions of DNA. 
Interestingly, we find that the Adams-Gibbs plot for the two different 
temperatures collapse on a single curve which can be fitted to a straight 
line. This seems to indicate that the activation energies for the transfer 
of water molecules between different regions of aqueous DNA are at most 
weakly dependent on temperature above the L-L transition. Dynamics below 
the L-L transition is too slow to allow a comprehensive study. The present
calculation of entropy is semi-quantitatively reliable as the entropy of
bulk water is correctly (within 5 \%) reproduced and also the chemical
potentials of bulk water and groove water are found to be the same, as
expected for systems in equilibrium.

\begin{figure}
\epsfig{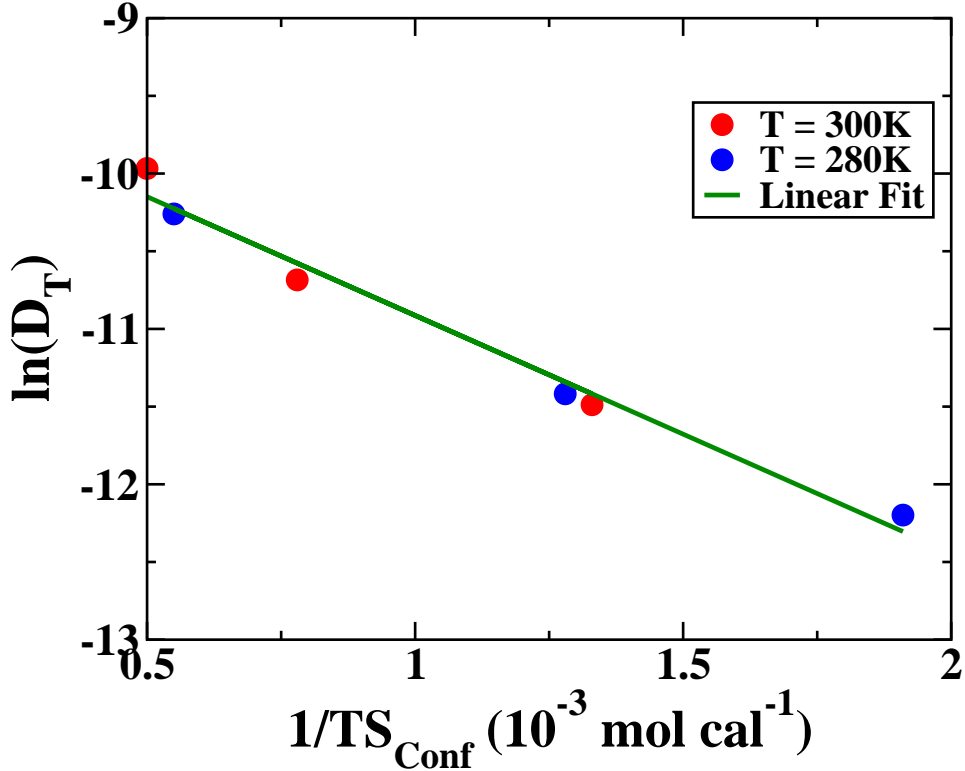}
\caption{ Adam-Gibb's plot of translational diffusivity (ln($D_T$)) vs
1/$TS_{Conf}$ for water molecules in the different regions of DNA
duplex (major groove, minor groove and bulk) at two different
temperatures, T = 300 K and T = 280 K.} 
\end{figure}

In a thermodynamic co-existence between two phases, a discontinuous change 
in entropy signals the presence of latent heat and a first order phase 
transition. However, in the present case, the large difference in entropy 
between bulk and minor groove water molecules should be regarded as a 
signature in the difference in structure between the two phases. Because of 
the small number ($\sim$65 for major groove and $\sim$30 for minor groove) 
of water molecules present in the groove region, a detailed quantification 
of microscopic structural arrangement is hard to perform.

Because the numbers of water molecules in the two grooves of the dodecamer 
are rather small, we have also simulated a large system with a 
standardized 38 base pair DNA and 8,000 water molecules interacting 
with TIP3P potential~\cite{maiti_goddard_nar04_bj06}. This system is known 
to sustain a stable double helix over a long time 
period~\cite{maiti_goddard_nar04_bj06}. Interestingly, we obtained 
qualitatively similar results for the groove water molecules, but 
transitions (around 245 K in the grooves) are not as prominent since TIP3P 
is known not to be a good network forming liquid and the {\em L-L transition 
is largely suppressed in the bulk phase}. Nevertheless, we do find similar 
kind of transition in the grooves in two different systems with two 
different water models which strengthens the generality of the results 
obtained in the present study.

\section{Conclusions}
In conclusion, we have studied both translational and rotational motions 
of water in the grooves of a  DNA duplex . We find that groove water shows 
a remarkable dynamical transition which can explain the transition observed 
in DNA duplex itself. {\em The fact that this transition occurs at not too 
deeply supercooled water ($T_{GL}\approx$ 255 K) suggests that this can 
be of importance in natural world}.

\section{Acknowledgement}
We thank Dr. Sarika Maitra Bhattacharyya and Prof. Srikanth Sastry for 
useful discussions. This work was supported in parts by a grant 
from DST, India. B.B. acknowledges support from J. C. Bose Fellowship. 
BJ and SP thank CSIR, India for providing SRF.

\end{document}